\documentclass[a4paper,12pt]{article}
\usepackage{amsmath}
\usepackage{graphicx}
\usepackage{amsfonts}

\title{Multi-instanton and String Loop Corrections in Toroidal
Orbifold Models}
\begin{document}

\thispagestyle{empty}

\vspace{-2cm}

\begin{flushright}
{\small CPHT-RR040.0608 \\
LPT-ORSAY 08-57}
\end{flushright}
\vspace{1cm}

\begin{center}
{\bf\LARGE Multi-instanton and String Loop Corrections in Toroidal
Orbifold Models}

\vspace{0.8cm}

{\bf Pablo G. C\'amara}$^{+}$ {\bf,\hspace{.2cm} Emilian
Dudas}$^{+,\ \dag}$ \vspace{0.8cm}

{\it $^{+}$ Centre de Physique Th\'eorique,\footnote{Unit\'e mixte
du CNRS, UMR 7644.}
Ecole Polytechnique,\\
F-91128 Palaiseau, France.\\[24pt]
$^{\dag}$ LPT,\footnote{Unit\'e mixte du CNRS, UMR 8627.}
Bat. 210, Univ. de Paris-Sud,\\
F-91405 Orsay, France.}

\vspace{0.7cm}

{\bf Abstract}
\end{center}
\vspace{.2cm} We analyze $\mathcal{N}=2$ (perturbative and
non-perturbative) corrections to the effective theory in type I
orbifold models where a dual heterotic description is available.
These corrections may play an important role in phenomenological
scenarios. More precisely, we consider two particular
compactifications: the Bianchi-Sagnotti-Gimon-Polchinski orbifold
and a freely-acting $\mathbb{Z}_2\times \mathbb{Z}_2$ orbifold
with  $\mathcal{N}=1$ supersymmetry and gauge group $SO(q)\times
SO(32-q)$. By exploiting
perturbative calculations of the physical gauge couplings on the
heterotic side, we obtain multi-instanton and one-loop string
corrections to the K\"ahler potential and the gauge kinetic
function for these models. The non-perturbative corrections appear
as sums over relevant Hecke operators, whereas the one-loop
correction to the K\"ahler potential matches the expression
proposed in \cite{fabre,haack}. We argue that these corrections
are universal in a given class of models where target-space
modular invariance (or a subgroup of it) holds. \clearpage

\tableofcontents

\section{Introduction}

In the last years there has been a remarkable progress in
understanding the structure of String Theory at tree-level in the
perturbative expansion, that is, in the supergravity limit. Flux
compactifications \cite{flux} have provided us with a helpful
framework which partially addresses long-standing problems such as
moduli stabilization or supersymmetry breaking. However, despite
this progress, the resulting message continues to be that
non-perturbative and string loop corrections play an indispensable
role, mainly due to the generic presence of remnant flat
directions in the scalar potential and the difficulties of
obtaining a chiral spectrum in their absence.

The computation of $\alpha'$ and non-perturbative corrections to
the effective theory is still in an early stage, even for cases
where a description in terms of a free CFT is available. Much
effort has been pursued on understanding the role of
$\mathcal{N}=1$ euclidean brane instantons \cite{instant1}. These
may turn out to be useful for generating new couplings in the
superpotential \cite{pertur1,pertur2,pertur3,pertur4,pertur5},
moduli-stabilization \cite{moduli} or supersymmetry breaking
\cite{susybreak}. Moreover, the one-loop string corrections to the
K\"ahler potential computed in \cite{alpha1}, have been shown to
play an important role in large volume scenarios \cite{large},
leading to a hierarchy of mass scales without the necessity of a
big amount of fine-tuning. Additional one-loop string corrections
to the K\"ahler potential have been computed in \cite{fabre,haack}
for toroidal compactifications (see also \cite{jump,alpha2}).

In this note, we analyze multi-instanton and one-loop string
corrections arising from $\mathcal{N}=2$ sectors in toroidal
orbifold compactifications of type I String Theory. These
generically correct the K\"ahler potential, $K$, and the gauge
kinetic function, $f_a$, of the effective theory, and therefore
may play an important role in phenomenological scenarios with
classical flat directions. For this aim, we follow the techniques
introduced in \cite{freelyact}, and build type I-heterotic S-dual
pairs of orbifold models.

Schematically, the procedure can be summarized as follows. The
one-loop physical gauge couplings in the heterotic side take the
expression,
\begin{equation}
4\pi^2g_a^{-2}(\mu^2)|_{\rm
1-loop}=\frac{k_a}{\ell}+\frac{b_a}{4}\textrm{log}\frac{M_{\rm
s}^2}{\mu^2}+\frac{\Delta_a(M,\bar M)}{4}\ ,\label{thresgeneral}
\end{equation}
with $\ell$ the linear multiplet associated to the dilaton, $M_s$
the string scale, $M$ the moduli of the compactification and $k_a$
the normalization of the gauge group generators, determined by the
level of the corresponding Kac-Moody algebra. The $\beta$-function
coefficient, $b_a$, is given in terms of the quadratic Casimir
invariants of the gauge group,
\begin{equation}
b_a=\sum_r n_rT_a(r)-3T_a({\rm adj}_a)\ ,
\end{equation}
with $n_r$ the number of matter multiplets in the representation
$r$.

On the other hand, the field theory result reads \cite{kl1,kl},
\begin{multline}
4\pi^2g_a^{-2}(\mu^2)|_{\rm 1-loop}=\textrm{Re }f_a(M)|_{\rm
1-loop}+\frac{b_a}{4}\left(\textrm{log}\frac{M_{\rm
Planck}^2}{\mu^2}-\textrm{log}(S+\bar S)\right)+\\
+\frac{1}{4}\left(c_a\hat K(M,\bar M)-2\sum_rT_a(r)\textrm{log det
}Z_r(M,\bar M)\right)\ , \label{field}
\end{multline}
where det $Z_r$ is the determinant of the tree-level K\"ahler
metric associated to the matter multiplets in the representation
$r$, $\hat K(M,\bar M)$ the tree-level K\"ahler potential for the
moduli $M$ and,
\begin{equation}
c_a=\sum_rn_rT_a(r)-T({\rm adj}_a)\ .
\end{equation}
In order to compare (\ref{thresgeneral}) and (\ref{field}), it is
convenient to express the relation between the usual complex
axiodilaton $S$ and the linear multiplet $\ell$ as,
\begin{equation}
\frac{1}{\ell}=\textrm{Re }S-\frac{1}{4}\Delta_{\textrm{univ.}}
\end{equation}
with $\Delta_{\textrm{univ.}}$ a gauge group independent
(``universal'') function. In what follows we split
$\Delta_{\textrm{univ.}}$ in its harmonic and non-harmonic parts,
\begin{equation}
\Delta_{\textrm{univ.}}(M,\bar M)=V_{(1)}(M,\bar M)+H(M)+H^*(\bar
M)\ .
\end{equation}
In terms of these, the K\"ahler potential and the gauge-kinetic
function of the $\mathcal{N}=1$ effective theory are given to
one-loop by \cite{derending,kl,stiebergerii}\footnote{We have
defined $S$ in such a way that the harmonic part of the universal
threshold, $H(M)$, naturally corrects the holomorphic gauge
kinetic function. Notice that this is always possible since the
chiral field $S$ has no fixed relation to vertices of string
theory.},
\begin{align}
K|_{\rm 1-loop}&=-\textrm{log}\left(S+\bar
S-\frac12V_{(1)}(M,\bar M)\right)+\hat K(M,\bar M)\ ,\label{k1}\\
\textrm{Re }f_a|_{\rm 1-loop}&=k_a \textrm{Re }S +
\frac{1}{4}\left(\Delta_a(M,\bar M) -
V_{(1)}(M,\bar M)- c_a\hat K(M,\bar M)\nonumber\right. \\
&\left.\qquad \qquad \qquad \qquad \qquad \qquad
-2\sum_rT_a(r)\textrm{log det }Z_r(M,\bar M)\right)\label{k2}
\end{align}
We can then reinterpret the results in terms of $E1$
multi-instanton and string loop corrections on the type I side, as
$K$ and $f_a$ should be invariant (up to K\"ahler transformations)
under S-duality transformations.

In the present work, we consider in detail two classes of models
on which the heterotic S-dual partition function can be easily
worked out: the Bianchi-Sagnotti-Gimon-Polchinsky (BSGP) orbifold
\cite{gp1,gp2}, with gauge group $U(16)\times U(16)$, and the
$\mathbb{Z}_2\times \mathbb{Z}_2$ freely-acting orbifold model
with gauge group $SO(q)\times SO(32-q)$, presented in
\cite{freelyact} and based on the model of \cite{antoniadis} (see
also \cite{blumen2}). The motivation is multiple. First, based on
the modular symmetries preserved by the scalars in Calabi-Yau
compactifications, and more precisely on the axionic shift
symmetries, it has been argued in \cite{blumen} that
non-perturbative corrections to the gauge kinetic function should
appear in an exponentiated form, $\textrm{exp}(2\pi M)$, with $M$
the corresponding moduli. Our results on the BSGP orbifold show
indeed that global symmetries constrains very much the shape of
multi-instanton and string-loop corrections and, with few extra
assumptions, they are completely determined in models on which
modular invariance in the moduli space of the compactification
applies\footnote{For previous works on the role of global
symmetries for determining non-perturbative and string loop
corrections, see \cite{elias0,pioline}.}.

The freely-acting $SO(q)\times SO(32-q)$ orbifold model
represents, on the other hand, an example on which part of the
original modular symmetry is broken by the compactification. A
remarkable fact, already pointed out in \cite{freelyact}, is that
$E1$ instantons in this model always appear within multiplets
under the orbifold action (doublets or quadruplets). We believe
that this may be a general feature of flux compactifications,
where the fluxes gauge some of the original symmetries and induce
non-trivial discrete torsion. It is therefore interesting to see
how non-perturbative and loop corrections are affected by the
``background'' in this simple example. Moreover, it was also
observed in \cite{freelyact} that, in the heterotic S-duals, the
orbifold action on the winding modes was different for $q=0$ (mod
8) or $q=4$ (mod 4), pointing out a possible dependence of the
type I instantonic effects on the rank of the gauge group. It is
also our aim to make this dependence explicit.

Although similar computations to the ones performed here have been
carried out e.g. for $R^4$ \cite{elias0,stiebergeriv}, $F^4$
\cite{elias1,stiebergeri,stiebergeriv,elias2} and the four
hyperini \cite{bianchi} couplings, this is to our knowledge the
first explicit computation of stringy multi-instanton corrections
to the K\"ahler potential and the gauge kinetic function. We hope
that these results will help to shed some light on some of the
issues raised in the previous paragraphs, and more interestingly,
to clarify the possible role of these corrections in
phenomenological scenarios.

The paper is organized as follows. In section \ref{sec2} we
construct the partition function for the heterotic S-dual of the
BSGP orbifold model, and extract the $E1$ multi-instanton and
one-loop string corrections to the K\"ahler potential and the
gauge kinetic function in the resulting effective theory. In
section \ref{sec3} we proceed similarly with the $SO(q)\times
SO(32-q)$ freely acting orbifold model. We comment on the possible
``universality'' of some of our results and discuss about possible
generalizations in section \ref{final}. Finally, we give some
concluding remarks in section \ref{final2}. We have relegated all
the details on the computations to the appendix, in order not to
overload the bulk of the paper with many technicalities.


\section{The Bianchi-Sagnotti-Gimon-Polchinski orbifold}
\label{sec2}

In this section we consider the BSGP type I orbifold model
\cite{gp1,gp2}, corresponding to the $T^4/\mathbb{Z}_2\times T^2$
orbifold limit of type I String Theory compactified in $K3\times
T^2$. In order to cancel the RR tadpoles, 8 D5-branes and 16
D9-branes are required. For D5 branes lying on top of an orbifold
fixed point, the complete massless spectrum has a $U(16)\times
U(16)$ gauge group with hypermultiplets in
$2(\mathbf{120},\mathbf{1})+2(\mathbf{1},\mathbf{120})+(\mathbf{16},\mathbf{16})$.
In the Coulomb branch, where half D5-brane is located at each of
the 16 fixed points, the Green-Schwarz mechanism takes place and
only the $U(16)$ gauge group from the D9-branes remains massless,
with spectrum given by four hypermultiplets, containing the moduli
of the $K3$, three vector multiplets containing the axiodilaton
and the moduli of the $T^2$, a
$\mathbf{120}+\overline{\mathbf{120}}$, and sixteen $\mathbf{16}$
coming from the D5-D9 modes. The coefficient of the
$\beta$-function turns out to be $b_{U(16)}=12$. Perturbative
threshold corrections to gauge couplings \cite{bf} depend on the
moduli of $T^2$, denoted $T_1$ and $U_1$ in what follows. Since
the dilaton $S$ and $T_1,U_1$ are in ${\cal N}=2$ vector
multiplets in 4d language, this is consistent with supersymmetry.
A priori we expect non-perturbative corrections to depend
nontrivially on the three vector multiplets, and to be insensitive
to the $T^4/\mathbb{Z}_2$ moduli, called $T_{2,3}$ and $U_{2,3}$
in what follows. We will show, by performing explicitly the
computation using the heterotic S-dual, that this expectation is
indeed correct.


\subsection{Heterotic S-dual partition function}

We want to find the one-loop partition function for the heterotic
dual of the BSGP model, proposed in \cite{polcho} for the above
Coulomb branch. In \cite{luis} it was shown that this corresponds
to a standard $SO(32)$ heterotic $T^2\times T^4/\mathbb{Z}_2$
orbifold with shift vector $V=\frac14(1,\ldots,1,-3)$. The various
orbifold blocks are then as follows. The left-moving fermions
contribute as\footnote{For definitions of the various modular
functions, affine characters and orbifold blocks, see e.g.
\cite{sagnot,libro}.},
\begin{equation}
Z_{\textrm{L}}\left[{h \atop
g}\right]=\frac{1}{2\eta^4}\sum_{a,b=0}^1(-1)^{a+b+ab+bh}
\vartheta^2\left[{a \atop b}\right]\vartheta^2\left[{a+h \atop
b+g}\right]\ ,
\end{equation}
with $h,g=0,1$ labelling the different untwisted and twisted
orbifold sectors.

\noindent Analogously, the bosonic $T^4$ blocks read,
\begin{equation}
Z_{(4,4)}\left[{0 \atop 0}\right]=\frac{\hat Z_2\hat
Z_3}{|\eta|^8}\quad \textrm{and} \quad Z_{(4,4)}\left[{h \atop
g}\right]=\left|\frac{2\eta}{\vartheta\left[{1-h \atop
1-g}\right]}\right|^4\ , \quad \textrm{for } hg\neq 0\ .
\end{equation}
Here we have defined the toroidal lattice sums $\hat Z_r$ as,
\begin{equation}
\hat Z_r=\frac{\textrm{Re
}T_r}{\tau_2}\sum_{n_1,\ell_1,n_2,\ell_2} \textrm{exp}\left[-2\pi
T_r \textrm{det}(A)-\frac{\pi (\textrm{Re }T_r)}{\tau_2(\textrm{Re
}U_r)}\left|\begin{pmatrix}1&iU_r\end{pmatrix}A\begin{pmatrix}\tau\\
-1\end{pmatrix}\right|^2\right]\ ,\label{latsum}
\end{equation}
\begin{equation}
A=\begin{pmatrix}n_1&\ell_1\\
n_2&\ell_2
\end{pmatrix}\ , \label{mat2}
\end{equation}
with $n_i$ and $\ell_i$ integers. For the right moving fermions we
find,
\begin{equation}
\Gamma\left[{h\atop
g}\right]=\frac{1}{2\bar\eta^{16}}\sum_{a,b=0}^1(-1)^{ga+hb}e^{-\frac{i\pi
hg}{2}}\bar\vartheta^{16}\left[{a-\frac{h}{2} \atop
b-\frac{g}{2}}\right] \ .
\end{equation}
Putting everything together we finally get the one-loop partition
function for the heterotic dual of the BSGP model,
\begin{multline}
T=\int_{\mathcal{F}} \frac{d^2\tau}{\tau_2^3}\frac{\hat
Z_1}{4|\eta|^8}\left[(Q_o+Q_v)\frac{\hat Z_2 \hat
Z_3}{|\eta|^8}\Gamma\left[{0 \atop 0}\right] +
(Q_o-Q_v)\left|\frac{2\eta}{\vartheta_2}\right|^4\Gamma\left[{0
\atop 1}\right]\right. +
\\
\left.
+(Q_s+Q_c)\left|\frac{2\eta}{\vartheta_4}\right|^4\Gamma\left[{1
\atop
0}\right]+(Q_s-Q_c)\left|\frac{2\eta}{\vartheta_3}\right|^4\Gamma\left[{1
\atop 1}\right]\right]\ , \label{partgp}
\end{multline}
where,
\begin{align}
Q_o+Q_v&=\frac{1}{2\eta^4}(\vartheta_3^4-\vartheta_1^4-\vartheta_2^4-\vartheta_4^4)\
, \\
Q_o-Q_v&=\frac{1}{2\eta^4}(\vartheta_3^2\vartheta_4^2-\vartheta_4^2\vartheta_3^2-\vartheta_1^2\vartheta_2^2-\vartheta_2^2\vartheta_1^2)\
, \end{align}\begin{align}
Q_s-Q_c&=\frac{1}{2\eta^4}(\vartheta_1^2\vartheta_3^2+\vartheta_3^2\vartheta_1^2+\vartheta_4^2\vartheta_2^2-\vartheta_2^2\vartheta_4^2)\
, \\
Q_s+Q_c&=\frac{1}{2\eta^4}(\vartheta_3^2\vartheta_2^2-\vartheta_2^2\vartheta_3^2+\vartheta_4^2\vartheta_1^2+\vartheta_1^2\vartheta_4^2)\
.
\end{align}

\subsection{Perturbative and non-perturbative corrections}
\label{gpth}

Following the general discussion around (\ref{thresgeneral}), our
task here is to compute the one-loop threshold corrections to the
physical gauge coupling in the heterotic model (\ref{partgp}), as
these are mapped to one-loop and $E1$ multi-instanton corrections
in the BSGP orbifold. In terms of the partition function, these
are given by \cite{kaplu,dkl,kk2},
\begin{equation}
\Lambda\equiv\frac{b_{U(16)}}{4}\textrm{log}\frac{M_{\rm
s}^2}{\mu^2}+\frac{\Delta_{U(16)}}{4}
=\int_{\mathcal{F}}\frac{d^2\tau}{\tau_2}\frac{i}{4\pi}\frac{1}{|\eta|^4}\sum_{a
  ,b=0}^1
\partial_\tau\left(\frac{\vartheta \left[{a \atop b }\right]}{\eta}\right)
\left(Q^2-\frac{1}{4\pi\tau_2}\right)C\left[{a \atop b }\right]\ ,
\label{threskap}
\end{equation}
where $Q$ is the charge operator of the corresponding gauge group,
and $C\left[{a \atop b }\right]$ is the internal six-dimensional
partition function. Following the same procedure than in
\cite{freelyact} we find,
\begin{equation}
\Lambda=-\frac18 \int_{\mathcal{F}}\frac{d^2\tau}{\tau_2} \hat
Z_1\hat{\mathcal{A}}_f\label{int}
\end{equation}
with,\footnote{The modular covariant derivative $D_d$ is defined
as,$$D_d=\frac{i}{\pi}\partial_{\tau}+\frac{d/2}{\pi \tau_2} $$}
\begin{equation}
\hat{\mathcal{A}}_f=-\frac{1}{20 \eta^{24}}(D_{10} E_{10} - 48
\eta^{24})=\frac{1}{12\eta^{24}}\left(\hat{E}_2 E_4
E_6-\frac{5}{12} E_6^2-\frac{7}{12} E_4^3\right)\ , \label{elip1}
\end{equation}
and $\hat Z_1$ given in (\ref{latsum}). The definitions of the
Eisenstein series, $E_{2k}$, can be found for instance in the
appendix of \cite{freelyact}.

The details of the computation are in appendix \ref{ap1}. Notice
that the numerator of $\hat{\mathcal{A}_f}$ is an
almost-holomorphic modular form, their non-holomorphicity being
exclusively due to the presence of $\hat E_2$,
\begin{equation}
\hat E_2\equiv E_2-\frac{3}{\pi \tau_2}\ .
\end{equation}
As it will be made more explicit below, these non-holomorphic
terms can be traced back to perturbative and non-perturbative
corrections to the K\"ahler potential of the effective theory.

Both, $\hat Z_1$ and $\hat{\mathcal{A}}_f$, are invariant under
the full modular group $\Gamma$, so we can directly apply the
method of Dixon-Kaplunovsky-Louis (DKL) \cite{dkl} to evaluate the
integral in (\ref{int}). This consists on depicting the lattice
sum, $\hat Z_1$, into orbits under the modular group, and evaluate
the integral for each class of orbits in a suitable unfolded
region of the upper complex half-plane. The matrices (\ref{mat2})
can be classified in three kind of orbits under the modular group,
\begin{enumerate}
\item \underline{Zero orbit:}
\begin{equation*}\begin{pmatrix}0& 0\\ 0& 0\end{pmatrix}\ ,\end{equation*}
\item \underline{Non-degenerate orbits:}
\begin{equation*}\begin{pmatrix}k& j\\ 0& p\end{pmatrix}\ ,\end{equation*} with
$k>j\geq 0$, $p\neq 0$ and $AV= AV'$ iff $V=V'$, for
$V,V'\in\Gamma$. \item \underline{Degenerate orbits:}
\begin{equation*}\begin{pmatrix}0& j\\ 0& p\end{pmatrix}\
,\end{equation*} with $(j,p)\sim(-j,-p)$ and $AV= AV'$ iff
$V=T^nV'$, for some integer $n$ and $V,V'\in\Gamma$.
\end{enumerate}
We therefore unfold (\ref{int}) into three integrals corresponding
to the above representatives. Non-degenerate orbits are integrated
over the double cover of the upper half complex plane,
$\mathbb{C}^+$, whereas degenerate orbits have to be integrated
over the fundamental domain, $\mathcal{F}_T$, of the subgroup
generated by $T$, for arbitrary $j$ and $p$. The details of the
computation can be found in appendix \ref{ap1}. Putting all pieces
together and disregarding constant terms arising from the
regularization scheme, we obtain,
\begin{multline}
\Lambda=\frac{\pi}{2} \textrm{Re }T_1 \ - \
3\left(\textrm{log}|\eta(iU_1)|^4+\textrm{log}[(\textrm{Re
}U_1)(\textrm{Re }T_1)\mu^2]\right)\ -\ \frac{\pi}{3}\frac{E(iU_1,2)}{T_1+\bar T_1}-\\
-\frac14\left( \sum_{k>j\geq 0, p>0}\frac{1}{kp}e^{-2\pi pk
T_1}\left[\hat{\mathcal{A}}_f(\mathcal{U})+\frac{1}{\pi
kp}\frac{\hat{\mathcal{A}}_K(\mathcal{U})}{T_1+\bar T_1}\right]\ +
\ \textrm{c.c.}\right)\label{gplambda}
\end{multline}
where $E(U,k)$ is the non-holomorphic Eisenstein series of order
$k$, defined as
\begin{equation}
E(U,k)\equiv
\frac{1}{\zeta(2k)}\sum_{(j_1,j_2)\neq(0,0)}\frac{(\textrm{Im
}U)^k}{|j_1+j_2U|^{2k}}\ ,
\end{equation}
and $\hat{\mathcal{A}}_K$ the almost-holomorphic modular form,
\begin{equation}
\hat{\mathcal{A}}_K=
\frac{1}{12\eta^{24}}(\hat E_2 E_4 E_6+2E_6^2+3E_4^3)\
.\label{elip2}
\end{equation}

\begin{figure}[!ht]
\begin{center}
\includegraphics[width=9cm]{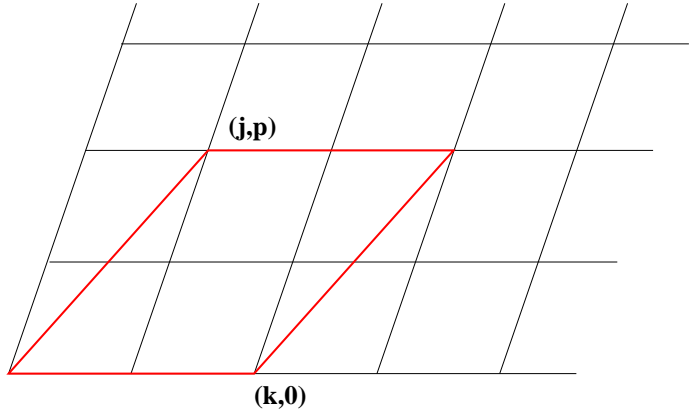}
\end{center}
\caption{\label{ff1}E1 multi-instanton wrapping the first 2-torus,
with induced worldvolume complex structure $\mathcal{U}$ given in
(\ref{uinst}).}
\end{figure}

The second term in (\ref{gplambda}) matches precisely the one-loop
threshold corrections computed in \cite{bf}, whereas the second
line in (\ref{gplambda}), corresponds to $E1$ multi-instanton
corrections. These are wrapping the first 2-torus, with induced
worldvolume complex structure \cite{elias1},
\begin{equation}
\mathcal{U}=\frac{j+ipU_1}{k}\ ,\label{uinst}
\end{equation}
as depicted in figure \ref{ff1}.
Their contribution can be also expressed as a sum over standard
Hecke operators acting on (almost-holomorphic) modular invariant
forms,
\begin{multline}
\sum_{k>j\geq 0, p>0}\frac{1}{kp}e^{-2\pi pk
T_1}\left[\hat{\mathcal{A}}_f(\mathcal{U})+\frac{1}{\pi
kp}\frac{\hat{\mathcal{A}}_K(\mathcal{U})}{T_1+\bar T_1}\right]=\\
\sum_{N=1}^\infty e^{-2\pi
NT_1}H_N\left[\hat{\mathcal{A}}_f+\frac{1}{N\pi}\frac{\hat{\mathcal{A}}_K}{T_1+\bar
T_1}\right](iU_1)\label{hecke}
\end{multline}
with,
\begin{equation}
H_N[\Phi](iU)=\frac{1}{N}\sum_{p>0,\ kp=N}\sum_{k>j\geq
0}\Phi\left(\frac{j+ipU}{k}\right)\ .
\end{equation}
It is thus evident the invariance of (\ref{gplambda}) under
$SL(2,\mathbb{Z})$ transformations of $U_1$, in agreement with the
global symmetry preserved by the orbifold.

In order to extract from (\ref{gplambda}) the corrections to the
effective theory, we need the K\"ahler metric for the D9-D9 and
D9-D5 matter fields. This is given by \cite{aspects,lust},
\begin{equation}
K_{C_k^{99}\bar C_k^{99}}=\frac{1}{(T_k+\bar T_k)(U_k+\bar U_k)}\
, \qquad K_{C_k^{95}\bar
C_k^{95}}=\prod_{j=2,3}\frac{1}{[(T_j+\bar T_j)(U_j+\bar
U_j)]^{1/2}}\ ,
\end{equation}
for $k=1,2,3$, so that,
\begin{multline}
\sum_rT_a(r)\textrm{log det }Z_r(M,\bar M) = \\
- 16\ \textrm{log}[(T_1+\bar T_1)(U_1+\bar U_1)] - 22
\sum_{j=2,3}\textrm{log}[(T_j+\bar T_j)(U_j+\bar U_j)] \ .
\end{multline}
From (\ref{k1}) and (\ref{k2}), then we read the following
expressions for the corrected K\"ahler potential and gauge kinetic
function in the effective theory,\footnote{We have performed an
expansion of the logarithm in eq.(\ref{k1}) around weak coupling.}
\begin{align}
&K=-\textrm{log}(S+\bar S)-\sum_{i=1}^3\textrm{log}[(T_i+\bar
T_i)(U_i+\bar U_i)]+\frac{1}{2}\frac{V_{1-loop}+V_{E1}}{S+\bar
S}\ , \\
&V_{1-loop}=-\frac{4\pi}{3}\frac{E(iU_1,2)}{T_1+\bar T_1}\ , \\
&V_{E1}=-\frac{1}{\pi}\sum_{k>j\geq 0,\ p>0}\frac{e^{-2\pi kp
T_1}}{(kp)^2}\left[\frac{\hat{\mathcal{A}}_K(\mathcal{U})}{T_1+\bar
T_1}-\frac{2i
kp}{\mathcal{U}-\bar{\mathcal{U}}}\frac{E_{10}(\mathcal{U})}{\eta^{24}(\mathcal{U})}\right]\
+ \ \textrm{c.c.}\ , \\
&f_{U(16)}=S+\frac{\pi T_1}{2}-12 \textrm{log
}\eta(iU_1)-\frac12\sum_{k>j\geq 0,\ p>0}\frac{e^{-2\pi
kpT_1}}{kp}\mathcal{A}_f(\mathcal{U})\ ,\label{gaug}
\end{align}
where the holomorphic modular form $\mathcal{A}_f$ is defined as
in (\ref{elip1}), replacing $\hat E_2$ by $E_2$. Several comments
are in order. First, observe that the $\beta$-function coefficient
exactly matches the field theory result. Moreover, the one-loop
$\alpha'$ correction to the K\"ahler potential agrees with the
expression obtained in \cite{fabre,haack} by direct computation in
the type I side. In our context, these corrections come from
non-holomorphic terms in the contributions of degenerate orbits.
Modular transformations of the $T_1$ modulus mix the $\alpha'$
corrections with the instantonic terms, in agreement with the fact
that T-duality is not a symmetry of type I String Theory. Notice
also that the loop correction of \cite{alpha1}, proportional to
$(\textrm{Re }S)^{3/2}$, is missing. This is consistent with the
fact that the internal torus has zero Euler characteristic,
$\chi=0$, for which the coefficient in front of the above
correction vanishes.

From the field theory perspective, the E1 multi-instanton
corrections of eq.(\ref{hecke}), enter as corrections to both the
K\"ahler potential and the holomorphic gauge kinetic function. To
our knowledge, these are new corrections and their role in the low
energy effective theory still has to be clarified. In section
\ref{final}, we will argue that these non-perturbative corrections
are general for any $\mathcal{N}=2$ sector in orbifold
compactifications where modular invariance of the target-space
holds.

Finally, the presence of the first term in the heterotic threshold
correction (\ref{gplambda}), contributing to the gauge kinetic
function (\ref{gaug}), may seem puzzling at first sight. Indeed,
by a straightforward counting of the string coupling, this linear
term in the $T^2$ volume modulus $T_1$, is expected to be a
tree-level (disk) effect on the type I side. On the other hand,
the $T_1$ modulus in the type I $\mathbb{Z}_2$ orbifold couples at
tree-level only to type I D5 branes. A possible origin is the
following. D9-branes in the BSGP model are fractional and
therefore its gauge kinetic function should receive a contribution
proportional to,
\begin{equation}
\sim \ \sum \sqrt{\textrm{det}(P[G+F_2])}\ T_1\ ,
\end{equation}
where the sum runs over the 16 singularities of $T^4/\mathbb{Z}_2$
and $P[\ldots]$ is the pull-back to the collapsed 2-cycle of the
singularity\footnote{We thank R. Blumenhagen for pointing out this
to us.}. In the orbifold limit, the volume of the 2-cycle is zero
and therefore the contribution from the metric vanishes. However,
as pointed out in \cite{polcho}, there is a non-trivial U(1) gauge
bundle on the collapsed 2-cycles which, in the blow-up limit,
leads together with the 8 D5-branes to the 24 instantons which are
required to satisfy RR 3-form Bianchi identity, $dF_3=\textrm{Tr
}R\wedge R-\textrm{Tr }F_2\wedge F_2$, in a smooth K3. It is
therefore expected a linear contribution to the gauge kinetic
function of the D9-brane from this hidden U(1) bundle at the
singularities.

\subsection{$E1$ instantons}
\label{e1}

Type I String Theory and its toroidal orbifolds has E5 instantons
wrapping the whole internal space and E1 instantons wrapping
various two cycles, in our case instantons E1$_i$ wrapping the
$T^2$ torus and various two cycles inside $T^4/Z_2$. Since the
instantonic corrections computed in the previous section depend on
the moduli of the $T^2$ torus, from the type I point of view they
should come from E1 instantons wrapping $T^2$. These instantons
are of two different types, depending if they sit or not at
$\mathbb{Z}_2$ orbifold fixed points.
\begin{itemize}
\item \emph{$E1$ instantons at orbifold fixed points.} These
instantons have unitary Chan-Paton factors, $U(r)$, with neutral
sector given by~:
\begin{itemize}
\item bosonic zero modes $x_{\mu}$, $y_{1,2}$ and fermionic zero
modes $\Theta^{\alpha,a}$, $\Theta^{\dot{\alpha},a}$, with $a=1,2$
in the adjoint representation ${\bf r} {\bf \bar r} $.  \item
bosonic zero modes $y_{3,4,5,6}$ and fermionic ones
$\lambda^{\alpha,a}$ in the symmetric representation $\frac{{\bf
r}({\bf r}+1)}{2} + \frac{{\bf \bar r}( {\bf \bar r}+1)}{2} $.
\item fermionic zero modes $\tilde{\lambda}^{\dot{\alpha},a}$ in
the antisymmetric representation $\frac{{\bf r}({\bf r}-1)}{2} +
\frac{{\bf \bar r}( {\bf \bar r}-1)}{2} $.\end{itemize} Regarding
the charged zero modes stretched between the instanton and the
corresponding 1/2 D5-brane stuck at the singularity, we obtain:
\begin{itemize}
\item bosonic zero modes $\mu^{1,2}$ from the R sector and
fermionic zero modes $\omega^{\alpha}$ from the NS sector, in the
representation ${\bf r}_{-1} + {\bf \bar r}_{1} $, where the
subscript denotes the $U(1)_5$ charge. \item bosonic zero modes
$\mu'_{1,2}$  in the representation ${\bf r}_{+1} + {\bf \bar
r}_{-1} $.
\end{itemize}
Finally, from the E1-D9 strings, there is a bosonic zero mode
$\nu$ in the representation ${\bf r {\bar n}} + {\bf \bar r {n}}$.

\item \emph{$E1$ instantons off the orbifold fixed points.} These
instantons have orthogonal Chan-Paton factors $SO(d)$. Here we
simply give their neutral sector~:
\begin{itemize} \item bosonic zero modes $x_{\mu}$, $y_{3,4,5,6}$ and fermionic
zero modes $\Theta^{\alpha,a}$, $\Theta^{\dot{\alpha},a}$ in the
representation $\frac{{\bf d} ({\bf d}+1)}{2} $. \item bosonic
zero modes $y_{1,2}$ and fermionic ones $\lambda^{\alpha,a}$,
$\lambda^{\dot{\alpha},a}$  in the representation $\frac{{\bf d}
({\bf d}-1)}{2} $.\end{itemize}
\end{itemize}

In order the instantons to contribute to the gauge kinetic
function, only four fermionic neutral zero modes should be
massless (corresponding to the ``goldstinos'') \cite{blumen}.
Therefore, most of the above zero modes should be lifted by
interactions. A possible qualitative picture is then the
following\footnote{We thank very much A. Uranga for suggesting
this picture to us and patient explanations.}. First, notice that
a $U(1)$ instanton on top of a singularity correspond to a
``gauge'' instanton for the U(1) gauge theory inside the
corresponding half D5-brane. These instantons are analogous to the
ones discussed in \cite{petersson}, with the extra fermionic zero
modes being lifted by couplings involving the
D5-branes\footnote{$\mathcal{N}=2$ gauge instantons in String
Theory orbifolds have been also extensively discussed in
\cite{billo2}.}. Therefore they should be responsible of the
1-instanton ($N=1$) contribution in eq.(\ref{hecke}). Notice
however that in this case there is a Higgs branch which consists
on moving the instanton out of the singularity, leading to a SO(1)
instanton (plus its image under the orbifold). In this limit, the
instanton has too many zero modes and does not correct the gauge
kinetic function. Similar situations where instantons only
contribute in a given locus of their moduli space have been
extensively discussed in \cite{angel}.

Hence, generically, for the $N$-instanton contribution in
eq.(\ref{hecke}), the moduli space of the multi-instanton contains
a subspace consisting on deformations of the instanton along the
$T^4/\mathbb{Z}_2$ directions. In a generic point of this space
the instanton gauge group is $SO(1)^{N}$, and the number of
fermionic zero modes is too high. However, in the special locus on
which all the components of the multi-instantons are on top of the
same singularity, the instanton gauge group is enhanced to $U(N)$
and only four zero modes survive, with the extra zero modes
presumably lifted by interactions with the D5-branes.

\section{The $SO(q)\times SO(32-q)$ freely-acting orbifold}
\label{sec3}

We consider now a slightly more complex class of models, given by
the $\mathbb{Z}_2\times\mathbb{Z}_2$ freely-acting orbifold with
gauge group $SO(q)\times SO(32-q)$ presented in \cite{freelyact}.
As already mentioned in the introduction, the motivation is two
fold. First, to understand how non-perturbative effects are
affected by the presence of a ``background'', breaking some of the
original global symmetries. Second, to make more explicit and shed
some light on the dependence of the E1 instantonic corrections on
the rank of the gauge group for this class of models, as it was
pointed out in \cite{freelyact}.

In the type I side, the orbifold action on the internal
coordinates is given by,
\begin{align}
&(x^1,x^2,x^3,x^4,x^5,x^6) \rightarrow (x^1+1/2,x^2,-x^3,-x^4,-x^5+1/2,-x^6) \ , \\
&(x^1,x^2,x^3,x^4,x^5,x^6) \rightarrow (-x^1+1/2,-x^2,x^3+1/2,x^4,-x^5,-x^6) \ , \\
&(x^1,x^2,x^3,x^4,x^5,x^6) \rightarrow
(-x^1,-x^2,-x^3+1/2,-x^4,x^5+1/2,x^6) \ .
\end{align}
The massless $\mathcal{N}=1$ spectrum can be read from the
partition function (see \cite{freelyact} for details) and contains
one chiral multiplet in the bifundamental representation,
$(\mathbf{q},\mathbf{32-q})$. The $\beta$-function coefficient for
the $SO(q)$ gauge group factor then reads,
\begin{equation}
b_{SO(q)}=4q-38\ .\label{bet}
\end{equation}
Due to the discrete shifts, modular invariance of the underlying
$(T^2)^3$ is broken to a subgroup of it. Moreover, the $E1$
instantons no longer appear as singlets under the orbifold action,
but rather as doublets or quadruplets \cite{freelyact}. This kind
of behavior is expected to be generic e.g. in flux
compactifications, where the fluxes gauge some of the originally
present symmetries and induce torsional cycles.\footnote{A simple
case are compactifications on solvmanifolds, corresponding to
freely-acting orbifolds of toroidal fibrations.}


\subsection{Heterotic S-dual partition function}

The partition function of the corresponding heterotic dual model
was worked out in \cite{freelyact}. The action of the orbifold on
the internal coordinates is given again by a
$\mathbb{Z}_2\times\mathbb{Z}_2$ action. We have summarized in
Table \ref{tablita} how each generator, $f$, $g$, $h$, acts on the
six internal coordinates and the gauge lattice.
\begin{table}[!h]
\begin{center}\begin{tabular}{|c||c|c|c|c|c|c||c|c|}
\hline generator & $x_1$ & $x_2$ & $x_3$ & $x_4$ & $x_5$ & $x_6$ & $SO(q)$ & $SO(32-q)$\\
\hline $g$ & $+$ & $+$ & $-$ & $-$ & $-$ & $-$ & $+$ & $+$ \\
\hline $f$ & $-$ & $-$ & $+$ & $+$ & $-$ & $-$ & $+$ & $-$ \\
\hline $h$ & $-$ & $-$ & $-$ & $-$ & $+$ & $+$ & $+$ & $-$ \\
\hline
\end{tabular}\end{center}
\caption{Orbifold action on the internal coordinates and on the
gauge degrees of freedom in the fermionic
formulation.}\label{tablita}
\end{table}
In addition, the action of each generator is accompanied by a
shift in the masses of the lattice states with
$(\textrm{momentum},\textrm{winding})=(m,n)$ according to,
\begin{equation}
(m,n)\ \xrightarrow{X} \ (m+s_X,n+s'_X)\ , \qquad X=f,g,h\ .
\end{equation}
Worldsheet modular invariance (or equivalently level-matching in
the twisted sectors) then requires \cite{freelyact},
\begin{align}
&q=0 \ \textrm{mod} \ 8\ \Rightarrow \
s_f=s_h=s_g=s'_f=s'_h=s'_g=1/2\ , \label{orbi}\\
&q=4 \ \textrm{mod} \ 8\ \Rightarrow \ s_f=s_h=s_g=s'_g=1/2\ ,
s'_f=s'_h=0\ .\nonumber
\end{align}
This is enough to completely determine the partition function. The
concrete expressions can be found in \cite{freelyact}.

Making use of changes of variables of the form,
\begin{equation}
\int_{\mathcal{F}}\frac{d^2\tau}{\tau_2^2}\mathcal{V}(\tau)=\int_{\mathcal{M}^{-1}(\mathcal{F})}\frac{d^2\tau}{\tau_2^2}\mathcal{V}(\mathcal{M}(\tau))\
,
\end{equation}
with $\mathcal{M}$ a modular transformation, it is easy to
reexpress the partition function in the more compact
form,\footnote{These changes of variables are of course not
unique. We could have equally chosen a different set of modular
transformations $\mathcal{M}$ (coset representatives), leading to
a different integrand and integration region.}
\begin{multline}
T=\int_{{\cal F}\oplus S({\cal F})\oplus ST^{-1}({\cal F})}
\frac{d^2 \tau}{\tau_2^3 |\eta|^8}\frac{1}{4} \{
\left[\frac{1}{3}(\tau_{oo}+\tau_{og}+\tau_{oh}+
\tau_{of})\frac{\hat Z_1\hat Z_2\hat Z_3}{|\eta|^4}+\right. \\
\left.+
(\tau_{oo}+\tau_{og}-\tau_{oh}-\tau_{of})(-1)^{m_1+n_1}\hat Z_1
\left|\frac{4\eta^2}{\vartheta_2^2}\right|^2 \right]
(\overline{\chi_o + \chi_v})+ \\
+\left[
(\tau_{oo}-\tau_{og}+\tau_{oh}-\tau_{of})(-1)^{m_3+n_3+\frac{qn_3}{4}}\hat
Z_3+\right.\\ \left.
+(\tau_{oo}-\tau_{og}-\tau_{oh}+\tau_{of})(-1)^{m_2+n_2+\frac{qn_2}{4}}\hat
Z_2 \right] \left|\frac{4\eta^2}{\vartheta_2^2}\right|^2
(\overline{\chi_o - \chi_v}) \}  \ , \label{het10}
\end{multline}
where the characters $\chi_o$ and $\chi_v$ are given in the
fermionic formulation of the gauge degrees of freedom by,
\begin{equation}
\chi_o \ = \  O_{32-q} O_q \ + \ C_{32-q} C_q \quad , \quad \chi_v
\ = \ V_{32-q} V_q \ + \ S_{32-q} S_q \ ,\label{het06}
\end{equation}
with $O_r$, $V_r$, $S_r$ and $C_r$ the standard $SO(r)$ affine
characters. The lattice sums with a sign insertion are given by,
\begin{multline}
(-1)^{m_1+h n_1}\hat Z_i=\frac{\textrm{Re
}T_i}{\tau_2}\sum_{n_1,\ell_1,n_2,\ell_2}(-1)^{h n_1
\ell_1}\\
\times \textrm{exp}\left[-2\pi T_i \textrm{det}(A)-\frac{\pi
(\textrm{Re }T_i)}{\tau_2(\textrm{Re
}U_i)}\left|\begin{pmatrix}1&iU_i\end{pmatrix}A\begin{pmatrix}\tau\\
-1\end{pmatrix}\right|^2\right]\ , \label{lat11}
\end{multline}
and,
\begin{equation}
A=\begin{pmatrix}n_1&\ell_1+\frac12\\
n_2&\ell_2
\end{pmatrix}\label{mat}
\end{equation}
The whole KK spectrum precisely matches the corresponding one on
the type I S-dual side, whereas the massive winding states and the
massive twisted spectra are, as expected, quite different. It
should be also noticed that while the  KK spectra are actually the
same for the two cases, $q=0$ and $q=4$ (mod $8$), they are very
different in the massive winding sector. We refer the interested
reader to \cite{freelyact} for the concrete expressions of the
partition functions in the type I S-dual side and other details.

\subsection{Perturbative and non-perturbative corrections}

\begin{figure}[!ht]
\begin{center}
\includegraphics[width=7cm]{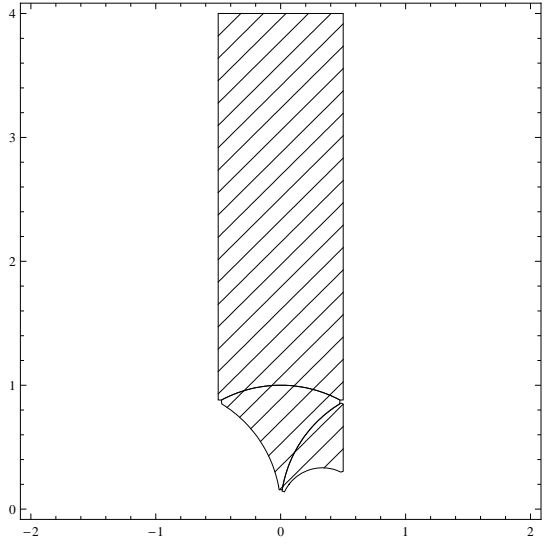}
\end{center}
\caption{\label{f1}Representation of ${\cal F}_{\Gamma_0(2)}$, the
fundamental domain for $\Gamma_0(2).$}
\end{figure}

Starting with the partition function (\ref{het10}) and proceeding
in the same way as we did with the BSGP orbifold, it can be shown
that the threshold corrections to the physical gauge couplings
(c.f. eq.(\ref{thresgeneral})) are given in this case by,
\begin{multline}
\Lambda_{SO(q)}\equiv\frac{b_{SO(q)}}{4}\textrm{log}\frac{M_{\rm
s}^2}{\mu^2}+\frac{\Delta_{SO(q)}}{4}=\\=-\frac{1}{4}\int_{{\cal
F}_{\Gamma_0(2)}} \frac{d^2 \tau}{\tau_2} \left[(-1)^{m_1+n_1}\hat
Z_1\hat{\mathcal{A}}_{f,1}^{[0,1]}+
\sum_{r=2,3}(-1)^{m_r+n_r+\frac{qn_r}{4}}\hat
Z_r\hat{\mathcal{A}}_{f,2}^{[0,1]}\right]\ ,\label{lab1}
\end{multline}
where,
\begin{align}
\hat{\mathcal{A}}_{f,1}^{[0,1]}(\tau)&=\frac{\vartheta^2_3
\vartheta^2_4 E_4( {\hat E}_2 E_4-
E_6)}{12\eta^{24}}\label{a1}\\
\hat{\mathcal{A}}_{f,2}^{[0,1]}(\tau)&=\frac{
\vartheta^2_3\vartheta^2_4}{24\eta^{24}}\left[
\vartheta_3^{q/2}\vartheta_4^{16-q/2}({\hat
E}_2+\vartheta_2^4-\vartheta_4^4)+
\vartheta_4^{q/2}\vartheta_3^{16-q/2}({\hat
E}_2-\vartheta_2^4-\vartheta_3^4)\right]\label{a2}
\end{align}
The details can be found in appendix \ref{ap2}. In order to
perform this integral, notice that the integration region,
\begin{equation}
\mathcal{F}_{\Gamma_0}(2)\equiv {\cal F}\oplus S({\cal F})\oplus
ST^{-1}({\cal F})\ ,
\end{equation}
which we have represented in figure \ref{f1}, corresponds to the
fundamental domain of the congruence subgroup $\Gamma_0(2)\subset
SL(2,\mathbb{Z})$. This consists of the modular matrices of the
form \cite{modular,modular2},
\begin{equation}
\begin{pmatrix}2a+1 & b\\ 2c & 2d+1\end{pmatrix}
\end{equation}

\begin{figure}[!ht]
\begin{center}
\includegraphics[width=15cm]{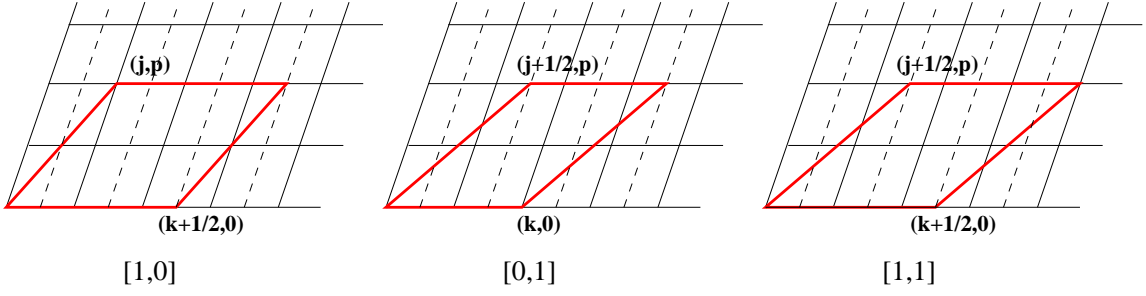}
\end{center}
\caption{\label{ff2}The three possible types of $E1_r$
multi-instantons, $[h,g]=\{[1,0], [0,1], [1,1]\}$, wrapping
torsional cycles in the $r$-th 2-torus in the $SO(q)\times
SO(32-q)$ model, with induced worldvolume complex structure
$\mathcal{U}_r^{[h,g]}$ given in eq.(\ref{udisc}). The continues
lines represent the lattice of the underlying 2-torus. The
orbifold generator reversing the transverse coordinates to the
instanton, shifts the lattice to the dashed one.}
\end{figure}

The generators of $\Gamma_0(2)$ are $T$ and $ST^2S$. Under these,
$\hat{\mathcal{A}}_{f,2}^{[0,1]}$ transforms as,
\begin{equation}
\label{tran} \hat{\mathcal{A}}_{f,2}^{[0,1]} \xrightarrow{T}
\hat{\mathcal{A}}_{f,2}^{[0,1]} \ , \quad
\hat{\mathcal{A}}_{f,2}^{[0,1]} \xrightarrow{ST^2S}
(-1)^{q/4}\hat{\mathcal{A}}_{f,2}^{[0,1]} \ ,
\end{equation}
whereas $\hat{\mathcal{A}}_{f,1}^{[0,1]}$ keeps invariant. We can
therefore classify the matrices (\ref{mat}) in orbits under
$\Gamma_0(2)$ in order to unfold the integral (\ref{lab1}),
similarly to what we did for the BSGP model.\footnote{One could
worry about the sign in the transformation of
$\hat{\mathcal{A}}_{f,2}^{[0,1]}$ under $ST^2S$, for $q=4$ mod 8.
However, this is automatically cancelled by the transformation of
the lattice sum,
\begin{equation}
(-1)^{m_r+n_r+\frac{qn_r}{4}}\hat Z_r\xrightarrow{ST^2S}
(-1)^{q/4}(-1)^{m_r+n_r+\frac{qn_r}{4}}\hat Z_r
\end{equation}
as required by modular invariance of (\ref{lab1}). Alteratively,
we could have performed an extra change of variables in
(\ref{lab1}) and reexpress it as an integral over the fundamental
domain of $\Gamma_0(4)\subset\Gamma_0(2)$, given by modular the
matrices of the form,
\begin{equation}
\begin{pmatrix}2a+1 & b\\ 4c & 2d+1\end{pmatrix}
\ ,\end{equation} obtaining the same final result.} There are four
kinds of orbits (three non-degenerate and one degenerate), whose
representatives can be taken to be,
\begin{enumerate}
\item \underline{Degenerate orbits:}
\begin{equation*}\begin{pmatrix}0& j+\frac12\\ 0& p\end{pmatrix}\ ,\end{equation*}
with $(j,p)\sim(-j-1,-p)$ and $AV= AV'$ iff $V=T^nV'$ for some
integer $n$ and $V,V'\in\Gamma_0(2)$. \item
\underline{Non-degenerate orbits:}
\begin{equation*}\textrm{I:} \quad \begin{pmatrix}k& j+\frac12\\ 0& p\end{pmatrix}\ ,
\qquad \textrm{II:} \quad \begin{pmatrix}j& -k-\frac12\\ p&
0\end{pmatrix}\ , \qquad \textrm{III:} \quad \begin{pmatrix}j-k&
-k-\frac12\\ p& 0\end{pmatrix}\ ,
\end{equation*}
with $k>j\geq 0$, $p\neq 0$ and $AV= AV'$ iff $V=V'$, for
$V,V'\in\Gamma_0(2)$.
\end{enumerate}
We can therefore unfold (\ref{lab1}) into four integrals
corresponding to the above representatives. The details are again
relegated to the appendix. Putting all pieces together we obtain,
\begin{multline}
\Lambda_{SO(q)}=-\frac{\pi}{2}\left[5E_{1/2}(iU_1,1)+\frac{q-17}{3}\sum_{r=2,3}E_{1/2}(iU_r,1)\right]+\\
+\frac{\pi}{360}\left[124\frac{E_{1/2}(iU_1,2)}{T_1+\bar
T_1}+\frac12\sum_{r=2,3}(q^2-32q+248)\frac{E_{1/2}(iU_r,2)}{T_r+\bar
T_r}\right]-\\
-\frac14 \sum_{k>j\geq 0,\
p>0}\sum_{[h,g]}\frac{(-1)^{gk+hj}}{pk_h}\left[e^{-2\pi
pk_hT_1}\left(\hat{\mathcal{A}}_{f,1}^{[h,g]}\left(\mathcal{U}_1^{[h,g]}\right)+\frac{1}{\pi
pk_h}\frac{\hat{\mathcal{A}}_{K,1}^{[h,g]}\left(\mathcal{U}_1^{[h,g]}\right)}{T_1+\bar
T_1}\right)+\right.\\
+\left.\sum_{r=2,3}(-1)^{(hj+gk)\frac{q}{4}}e^{-2\pi
pk_hT_r}\left(\hat{\mathcal{A}}_{f,2}^{[h,g]}\left(\mathcal{U}_r^{[h,g]}\right)+\frac{1}{\pi
pk_h}\frac{\hat{\mathcal{A}}_{K,2}^{[h,g]}\left(\mathcal{U}_r^{[h,g]}\right)}{T_r+\bar
T_r}\right)\right]\ + \ \textrm{c.c.}\ ,\label{so}
\end{multline}
where $\hat{\mathcal{A}}_{f,i}^{[h,g]}$ and
$\hat{\mathcal{A}}_{K,i}^{[h,g]}$, $i=1,2$, are given in appendix
\ref{elipo}, $k_h\equiv k+\frac{h}{2}$, and the shifted
non-holomorphic Eisenstein series, $E_{1/2}(U,k)$, are defined as,
\begin{equation}
E_{1/2}(U,k)\equiv
\frac{1}{\zeta(2k)}\sum_{j_1,j_2}\frac{(\textrm{Im
}U)^k}{|j_1+j_2U+1/2|^{2k}}\ , \label{shifted}
\end{equation}
In particular \cite{freelyact},
\begin{equation}
E_{1/2}(iU,1)=-\frac{3}{\pi}(\textrm{log}|\vartheta_2(iU)|^4+\pi\textrm{Re
}U+\textrm{log}[(\textrm{Re }U)(\textrm{Re }T)\mu^2])\ +\
\textrm{const.}\ .
\end{equation}
The sum on $[h,g]$ in (\ref{so}) extends over $[1,0]$, $[0,1]$ and
$[1,1]$, labelling the three types of $E1_r$ multi-instantons
contributing to (\ref{lab1}). The induced complex structure on
their worldvolume is given by,
\begin{equation}
\label{udisc} \mathcal{U}_r^{[h,g]}=\frac{j+ipU_r+g/2}{k+h/2}\ ,
\end{equation}
corresponding to instantons wrapping the torsional cycles of the
twisted cohomology, as illustrated in figure \ref{ff2}, or
alternatively, multi-instantons with discrete Wilson lines
$(\alpha,\beta)\in \{(0,\frac12),(1,0),(1,\frac12)\}$
\cite{blumen2}.

\vspace{0.5cm}

Subtracting the gauge group dependent part, along the lines of
eqs.(\ref{k1}) and (\ref{k2}), we obtain the following corrections
to the effective K\"ahler potential and gauge kinetic function,
\begin{align}
&K=-\textrm{log}(S+\bar S)-\sum_{i=1}^3\textrm{log}[(T_i+\bar
T_i)(U_i+\bar
U_i)]+\frac{1}{2}\sum_{i=1}^3\frac{V^i_{1-loop}+V^i_{E1}}{S+\bar
S}\ , \\
&V^1_{1-loop}=\frac{62\pi}{45}\frac{E_{1/2}(iU_1,2)}{T_1+\bar
T_1}\ ,\\
&V^r_{1-loop}=\frac{\pi}{180}(q^2-32q+248)\frac{E_{1/2}(iU_r,2)}{T_r+\bar
T_r}\ , \\ &V^1_{E1}=-\sum_{k>j\geq 0,\
p>0}\sum_{[h,g]}\frac{(-1)^{gk+hj}}{p^2k_h^2}e^{-2\pi pk_hT_1}\\
&\quad \times
\left[\frac{\hat{\mathcal{A}}_{K,1}^{[h,g]}\left(\mathcal{U}_1^{[h,g]}\right)}{T_1+\bar
T_1}-\frac{i\pi
pk_h}{\mathcal{U}_1^{[h,g]}-\bar{\mathcal{U}}_1^{[h,g]}}\chi_1\left[{h
\atop g}\right](\mathcal{U}_1^{[h,g]})\right]\ + \
\textrm{c.c.}\nonumber \\ &V^r_{E1}=-\sum_{k>j\geq 0,\
p>0}\sum_{[h,g]}\frac{(-1)^{(hj+gk)\left(\frac{q}{4}+1\right)}}{p^2k_h^2}e^{-2\pi
pk_hT_r}\\
&\quad
\times\left[\frac{\hat{\mathcal{A}}_{K,2}^{[h,g]}\left(\mathcal{U}_r^{[h,g]}\right)}{T_r+\bar
T_r}-\frac{i\pi
pk_h}{\mathcal{U}_r^{[h,g]}-\bar{\mathcal{U}}_r^{[h,g]}}\chi_2\left[{h\atop
g}\right](\mathcal{U}_r^{[h,g]})\right]\ + \ \textrm{c.c.}\
,\nonumber\\ &f_{SO(q)}=S+\frac{15\pi}{2}U_1+30\textrm{log
}\vartheta_2(iU_1)+(q-17)\sum_{r=2,3}\left[\frac{\pi}{2}U_r+2\textrm{log
}\vartheta_2(iU_r)\right]-\\&\qquad\qquad -\frac12 \sum_{k>j\geq
0,\ p>0}\sum_{[h,g]}\frac{(-1)^{gk+hj}}{pk_h}\left[e^{-2\pi
pk_hT_1}\mathcal{A}_{f,1}^{[h,g]}(\mathcal{U}_1^{[h,g]})+\right.\nonumber\\
&\left.\qquad\qquad\qquad\qquad\qquad\qquad
+\sum_{r=2,3}(-1)^{(hj+gk)\frac{q}{4}}e^{-2\pi
pk_hT_r}\mathcal{A}_{f,2}^{[h,g]}(\mathcal{U}_r^{[h,g]})\right]\
,\nonumber
\end{align}
where $\mathcal{A}_{f,i}^{[h,g]}$ is defined as
$\hat{\mathcal{A}}_{f,i}^{[h,g]}$, but replacing $\hat E_2$ by
$E_2$, and we have introduced the notation,
\begin{align*}
&\chi_1\left[{h \atop
g}\right](\tau)\equiv\frac{4(\chi_o+\chi_v)}{\eta^2\vartheta\left[{1-h
\atop 1-g}\right]^2}\ , & &\chi_2\left[{0 \atop
1}\right](\tau)\equiv
\frac{\chi_o-\chi_v}{\eta^8}\vartheta_3^2\vartheta_4^2 \ , \\
&\chi_2\left[{1 \atop 0}\right](\tau)\equiv \chi_2\left[{0 \atop
1}\right](S\tau)\ , & & \chi_2\left[{1 \atop 1}\right](\tau)\equiv
\chi_2\left[{0 \atop 1}\right](ST^{-1}\tau)\ .
\end{align*}
Several comments are in order. First, notice that the field theory
result for the $\beta$-function coefficient, (\ref{bet}), is
correctly reproduced. The overall structure of the
non-perturbative and loop corrections is very similar to the ones
in the BSGP orbifold, but the standard Eisenstein series and Hecke
operators are replaced by the corresponding automorphic forms of
$\Gamma_0(2)$. Moreover, there is a non-trivial dependence of the
non-perturbative dynamics on the rank of the gauge group, through
the phases $\textrm{exp}[i\pi(hj+gk)\frac{q}{4}]$. These would
explain why in the heterotic side the orbifold action on the
winding modes is very different, depending on the value of $q$
(c.f. eq.(\ref{orbi})). This behavior may resemble in spirit the
more familiar situation of ordinary gauge theory instantons, where
their contributions are often subjected to constraints depending
on the ranks of the gauge group.

By a direct inspection of (\ref{so}), it is easy to check the fact
that instantonic corrections are gauge-group independent when they
come from instantons which are left invariant by the orbifold
operations acting trivially on the gauge degrees of freedom;
whereas they are gauge-group dependent if they come from
instantons left invariant by the orbifold operations which act
non-trivially on the gauge degrees of freedom.

Finally, let us mention the possibility of additional
non-perturbative corrections coming from purely $\mathcal{N}=1$
sectors, not considered here. Precisely, in \cite{blumen2} it was
argued for the $q=32$ case, the presence of extra non-perturbative
contributions to the gauge thresholds, due to the combined effect
of multi-instantons wrapping different cycles of the internal
space.

\section{Universality of $\mathcal{N}=2$ corrections}
\label{final}

It has been pointed out very often in the literature the
importance of global symmetries in determining the expression of
non-perturbative and $\alpha'$ corrections which come from BPS
states \cite{pioline}. The results in the previous sections, based
on the S-duality map, reveal that the string loop and $E1$
multi-instanton effects coming from $\mathcal{N}=2$ subsectors of
the theory arise in terms of non-holomorphic Eisenstein series and
Hecke operators relevant to the global symmetry preserved by the
orbifold. In this section, we elaborate on certain
``universality'' of the $\mathcal{N}=2$ corrections computed in
the BSGP orbifold. Similar aspects have been discussed in the
context of $\alpha'$ corrections to the gauge couplings in
heterotic compactifications
\cite{dkl,eliasm,stiebergerii,stiebergeriii}.

Precisely, we would like to consider toroidal orbifold
compactifications on which the orbifold action, $\mathbb{G}$,
contains some subgroup, $\mathbb{G}^i$, leaving unrotated a given
complex plane. The contribution of these sectors to the threshold
corrections to the physical gauge couplings can be expressed as,
\begin{equation}
\Lambda_a=-\frac18\sum_i\int_{\mathcal{F}}\frac{d^2\tau}{\tau_2}\hat
Z_i\hat{\mathcal{A}}^a_{f,i}\ ,
\end{equation}
where the sum runs over the disjoint union of $\mathcal{N}=2$
subsectors, each leaving invariant a single complex plane, and the
gauge group is given by a product $G=\prod_a G_a$. The lattice
sums, $\hat{Z}_i$, are given in eq.(\ref{latsum}), where $T_i$ and
$U_i$ are now the K\"ahler and complex structure moduli of the
corresponding unrotated complex plane. Moreover,
$\hat{\mathcal{A}}^a_{f,i}\sim M^a_{i}/\eta^{24}$, with $M^a_i$ an
almost holomorphic modular form of degree 24.

The space of holomorphic forms of degree 24 is a vector space of
dimension 2, engendered by the Eisenstein series $E_6^2$ and
$E_4^3$ \cite{modular,modular2}. If we also allow for almost
holomorphic modular forms, we have to include in addition $\hat
E_2 E_{10}$.\footnote{We could also think about including terms
with higher powers of $\hat E_2$, e.g. $\hat E_2^2 E_4^2$.
However, these terms are forbidden by $\mathcal{N}=2$
supersymmetry \cite{elias1}.} Hence $M_i^a$ is in general
determined by three coefficients, which usually can be obtained
from the low energy spectrum. More precisely, imposing the absence
of tachyons in the spectrum, we obtain,
\begin{equation}
\hat{\mathcal{A}}^a_{f,i}=2b_i^a+\frac{\gamma_i}{20\eta^{24}}\left[D_{10}E_{10}-528\eta^{24}\right]\
,
\end{equation}
where $b_i^a$ is the $\beta$-function coefficient of the
$\mathcal{N}=2$ gauge theory associated to a would-be
$T^6/\mathbb{G}^i$ orbifold, $\gamma_i$ is a model dependent (but
gauge group independent) coefficient
to be determined, and we have made use of the identities,
\begin{equation}
D_{10}E_{10}=\frac23E_6^2+E_4^3-\frac53\hat E_2 E_{10}\ , \qquad
\eta^{24}=\frac{1}{2^6\cdot 3^3}[E_4^3-E_6^2]\ .
\end{equation}

Proceeding as in section \ref{gpth} we get,
\begin{multline}
\Lambda_a=\sum_i\bigg\{ \frac{\pi(b_i^a+6\gamma_i)}{12} \textrm{Re
}T_i \ + \
\frac{\pi\gamma_i}{3}\frac{E(iU_i,2)}{T_i+\bar T_i}\ -\\
-\frac14\left( \sum_{k>j\geq 0, p>0}\frac{1}{kp}e^{-2\pi pk
T_i}\left[\hat{\mathcal{A}}^a_{f,i}(\mathcal{U}_i)-\frac{\gamma_i}{\pi
kp}\frac{\hat{\mathcal{A}}_K(\mathcal{U}_i)}{T_i+\bar T_i}\right]\
+ \ \textrm{c.c.}\right)\ -\\
- \
\frac{b_i^a}{4}\left(\textrm{log}|\eta(iU_i)|^4+\textrm{log}[(\textrm{Re
}U_i)(\textrm{Re }T_i)\mu^2]\right)\bigg\}\ ,
\end{multline}
with $\hat{\mathcal{A}}_K$ and $\mathcal{U}_i$ defined in
(\ref{elip2}) and (\ref{uinst}), respectively. From this
expression, we can then extract the corrected K\"ahler potential
and gauge kinetic functions of the effective theory, as we did in
previous sections, obtaining
\begin{align}
&K=-\textrm{log}(S+\bar S)-\sum_{i}\bigg\{\textrm{log}[(T_i+\bar
T_i)(U_i+\bar U_i)]+\frac{1}{2}\frac{V^i_{1-loop}+V^i_{E1}}{S+\bar
S}\bigg\}+\ldots\ , \\
&V^i_{1-loop}=\frac{4\pi \gamma_i}{3}\frac{E(iU_i,2)}{T_i+\bar T_i}\ , \\
&V^i_{E1}=\frac{\gamma_i}{\pi}\sum_{k>j\geq 0,\ p>0}\frac{e^{-2\pi
kp
T_i}}{(kp)^2}\left[\frac{\hat{\mathcal{A}}_K(\mathcal{U}_i)}{T_i+\bar
T_i}-\frac{2i
kp}{\mathcal{U}_i-\bar{\mathcal{U}_i}}\frac{E_{10}(\mathcal{U}_i)}{\eta^{24}(\mathcal{U}_i)}\right]\
+ \ \textrm{c.c.}\ , \\
&f_a=S+\sum_i\bigg\{\frac{\pi
(b_i^{a}+6\gamma_i)T_i}{12}-b^a_i\textrm{log
}\eta(iU_i)-\frac12\sum_{k>j\geq 0,\ p>0}\frac{e^{-2\pi
kpT_i}}{kp}\mathcal{A}^a_{f,i}(\mathcal{U}_i)\bigg\}+\ldots \ ,
\end{align}
where the dots refer to possible additional corrections from other
sectors. The interpretation of these terms is similar to the one
discussed in sections \ref{gpth} and \ref{e1}.

Notice that these expressions in principle also apply in orbifolds
where the heterotic S-dual description is unknown, and therefore
our technique in principle no longer applies. It would be very
interesting to obtain the general formula for the $\mathcal{N}=2$
corrections by direct computation in the type I orbifold, and to
see whether there is agreement with our conjectured expression.

\section{Concluding remarks}
\label{final2}

In the present paper we explicitly computed the $E1$ instantonic
corrections to the gauge kinetic function $f$ and to the K\"ahler
potential $K$ in ${\cal N}=2$ and   ${\cal N}=1$ type I string
vacua which have known heterotic S-duals. We showed that one-loop
threshold corrections to gauge couplings in the heterotic dual
encode one-loop and instantonic corrections for both $f$ and $K$
on the type I side,  whereas the corresponding direct one-loop
type I threshold correction misses the one-loop correction to the
K\"ahler potential, computed by other methods in \cite{fabre} and
\cite{haack}. We gave arguments based on target-space modular
invariance on universality properties of instantonic corrections
in ${\cal N}=2$ vacua. It is clear however that our results apply
to the much larger class of models of ${\cal N}=1$ type I models
with ${\cal
 N}=2$ subsectors, like for example the $\mathbb{Z}_6$, $\mathbb{Z}_6'$, $\mathbb{Z}_8$ or $\mathbb{Z}_{12}$
type I orbifolds.

We performed a similar computation in dual pairs in
compactifications on smooth Calabi-Yau spaces which have an exact
CFT description,  based on a recently worked out class of
freely-acting S-dual pairs \cite{freelyact}. We showed that even
if the heterotic duals of perturbatively connected type I models
have different orbifold actions in the twisted (winding) sector,
the S-duality maps correctly heterotic $\alpha'$ corrections into
type I instantonic corrections. As a byproduct, we also checked
the intuitively obvious statement that instantonic corrections are
gauge-group independent if coming from instantons left invariant
by orbifold operations acting trivially on the gauge degrees of
freedom, whereas they are gauge-group dependent if coming from
instantons left invariant by orbifold operations acting
non-trivially on the gauge degrees of freedom.

As already argued in \cite{freelyact}, it is clear that whereas
our discussion was focused on multi-instantonic corrections to the
gauge kinetic function and the K\"ahler potential, similar
multi-instanton corrections are expected to occur for the
superpotential. A simple argument can be given in the case
(explicitly realized by the string construction of
\cite{freelyact}) where non-perturbative gauge (E5 instantonic)
effects occur on D9-branes, leading to a superpotential,
\begin{equation}
W_{np} \ = e ^{- b (S + f_1 (U_i) + f_{np}(U_i,T_i))} \ =
\ \sum_n c_n (U_i) \ e ^{- 2 \pi n T} \ e^{- b S} \ \label{gaugino}
\end{equation}

Whereas non-perturbative corrections to the superpotential are
well-known to play a crucial role in moduli stabilization
\cite{moduli}, we expect that instantonic corrections to the
K\"ahler potential may play also an important role in some
scenarios of moduli stabilization, for example in the large-volume
scenario \cite{large}. Moreover, the instantonic corrections to
the gauge kinetic function are expected to modify the gauge
couplings and gaugino masses, and in particular may become
relevant in concrete phenomenological models.

Another interesting direction which our paper has left partially
open is the detailed type I microscopic derivation of the
multi-instanton effects obtained here from S-duality, which should
involve in an important way the lifting of fermionic zero-modes by
instanton interactions along the lines of \cite{cvetic,angel}.

It would be, finally, very instructive to perform similar studies
in the S-dual pairs of $\mathcal{N}=1$ orbifold models conjectured
in \cite{n0,n1} and learn more about non-perturbative dynamics of
both sides using $\alpha'$ corrections on the heterotic side and
instantonic computations on the type I side.


\section*{Acknowledgments}
{We would like to thank C. Bachas, E. Kiritsis and M. Trapletti
for useful discussions and comments, and very especially R.
Blumenhagen, I. Garcia-Etxebarria and A. Uranga for very
illuminating comments on a draft version of the paper. This work
was supported by the ANR grant, ANR-05-BLAN-0079-02, the INTAS
contract 03-51-6346, the RTN contracts MRTN-CT-2004-005104 and
MRTN-CT-2004-503369, the CNRS PICS \#~2530, 3059 and 3747, the
MIUR-PRIN contract 2003-023852, the European Union Excellence
Grant MEXT-CT-2003-509661 and the NATO grant PST.CLG.978785.}


\appendix

\section{Details on the computations}

\subsection{The BSGP model}
\label{ap1}

\subsubsection{Elliptic genera}

In order to express the elliptic genus in terms of ordinary
modular forms we use the following relations,
\begin{align}
\vartheta^4\left[{0 \atop \pm 1/2}\right]&=\frac12
\vartheta_3\vartheta_4(\vartheta_3^2+\vartheta_4^2)\ , &
\vartheta^4\left[{1 \atop \pm
1/2}\right]&=\frac{\vartheta_2^3\eta^3}{\vartheta_3^2+\vartheta_4^2}
\ , \\
\vartheta^4\left[{\pm 1/2 \atop 0}\right]&=\frac12
\vartheta_2\vartheta_3(\vartheta_2^2+\vartheta_3^2)\ , &
\vartheta^4\left[{\pm 1/2 \atop
1}\right]&=-\frac{\vartheta_4^3\eta^3}{\vartheta_2^2+\vartheta_3^2}
\ , \\
\vartheta^4\left[{\pm 1/2 \atop \pm 1/2}\right]&=\frac12
\vartheta_2\vartheta_4(\vartheta_2^2-i\vartheta_4^2)\ , &
\vartheta^4\left[{\pm 1/2 \atop \mp
1/2}\right]&=\frac{\vartheta_3^3\eta^3}{\vartheta_2^2-i\vartheta_4^2}
\ ,
\end{align}
and,
\begin{align}
\frac{\vartheta''\left[{0 \atop \pm 1/2}\right]}{\vartheta\left[{0
\atop \pm
1/2}\right]}&=\frac{i\pi^3}{3}(4E_2+\vartheta_3^4-6\vartheta_3^2\vartheta_4^2+\vartheta_4^4)\
, \\
\frac{\vartheta''\left[{1 \atop \pm 1/2}\right]}{\vartheta\left[{1
\atop \pm
1/2}\right]}&=\frac{i\pi^3}{3}(4E_2+\vartheta_3^4+6\vartheta_3^2\vartheta_4^2+\vartheta_4^4)\
, \\
\frac{\vartheta''\left[{\pm 1/2 \atop
0}\right]}{\vartheta\left[{\pm 1/2 \atop
0}\right]}&=\frac{i\pi^3}{3}\left(4E_2+4\vartheta_3^4+\frac{(\vartheta_2^2-5\vartheta_3^2)\vartheta_4^4}{\vartheta_2^2+\vartheta_3^2}\right)\
, \\
\frac{\vartheta''\left[{\pm 1/2 \atop
1}\right]}{\vartheta\left[{\pm 1/2 \atop
1}\right]}&=\frac{i\pi^3}{3}\left(4E_2-8\vartheta_3^4+\frac{(\vartheta_2^2+7\vartheta_3^2)\vartheta_4^4}{\vartheta_2^2+\vartheta_3^2}\right)\
,\\ \frac{\vartheta''\left[{\pm 1/2 \atop \pm
1/2}\right]}{\vartheta\left[{\pm 1/2 \atop \pm
1/2}\right]}&=\frac{i\pi^3}{3}\left(4E_2-8\vartheta_4^4+7\vartheta_3^4-\frac{6\vartheta_2^2\vartheta_3^4}{\vartheta_2^2-i\vartheta_4^2}\right)\
, \\
\frac{\vartheta''\left[{\pm 1/2 \atop \mp
1/2}\right]}{\vartheta\left[{\pm 1/2 \atop \mp
1/2}\right]}&=\frac{i\pi^3}{3}\left(4E_2+4\vartheta_4^4-5\vartheta_3^4+\frac{6\vartheta_2^2\vartheta_3^4}{\vartheta_2^2-i\vartheta_4^2}\right)\
.
\end{align}

Then it is possible to show that,
\begin{align}
A_1&\equiv 2\left(\bar \Gamma\left[{0 \atop 1}\right]
\vartheta_3^2 \vartheta_4^2+ \bar \Gamma\left[{1 \atop 0}\right]
\vartheta_2^2 \vartheta_3^2+ \bar \Gamma\left[{1 \atop 1}\right]
\vartheta_2^2 \vartheta_4^2\right)=\frac{2 E_4E_6}{
\eta^{16}}=\frac{2 E_{10}}{ \eta^{16}}\ , \\
A_2&\equiv \partial^2_{\nu_1}A_1=-\frac{2\pi^2}{3 \eta^{16}}( E_2
E_4 E_6-\frac{5}{12} E_6^2-\frac{7}{12} E_4^3),
\end{align}
where $\partial_{\nu_1}$ acts on the first $SO(2)$ character in
$\bar \Gamma\left[{h \atop g}\right]$, with affine parameter
$\nu_1$. Therefore,
\begin{equation}
\hat{\mathcal{A}}_f\equiv
-\frac{1}{8\pi\bar\eta^8}\left(\frac{A_1}{\tau_2}+\frac{A_2}{\pi}\right)=\frac{1}{12\eta^{24}}(\hat{E}_2E_{10}-\frac{5}{12}
E_6^2-\frac{7}{12}
E_4^3)=-24+\frac{60}{\pi\tau_2}+\ldots\label{cala1}
\end{equation}
The other modular form that we will need in the computation of the
thresholds is,
\begin{equation}
\hat{\mathcal{A}}_K\equiv
\frac{1}{4\pi}\left(i\partial_\tau-\frac{1}{\tau_2}\right)\frac{E_{10}}{\eta^{24}}\label{cala2}
\end{equation}
Taking into account that $E_{10}=E_4E_6$ and,
\begin{equation}
\partial_\tau E_4=-\frac{2\pi i}{3}(E_6-E_2E_4)\ , \qquad
\partial_\tau E_6=-\pi i(E_4^2-E_2E_6)\ ,
\end{equation}
we then obtain,
\begin{equation}
\hat{\mathcal{A}}_K=\frac{1}{12\eta^{24}}(\hat E_2 E_4
E_6+2E_6^2+3E_4^3)\ .
\end{equation}

\subsubsection{Zero orbit}

For the zero orbit we have the contribution,
\begin{multline}
\Lambda_0=-\frac{\textrm{Re }T_1}{8}
\int_{\mathcal{F}}\frac{d^2\tau}{\tau_2^2}\hat{\mathcal{A}}_f=\\
=-\frac{\textrm{Re
}T_1}{96}\int_{\mathcal{F}}\frac{d^2\tau}{\tau_2^2}\left[\hat
E_2(e^{-2\pi i\tau}-240+\ldots)-e^{-2\pi i\tau}-24+\ldots\right]
\end{multline}
Making use of the formula \cite{lerche},
\begin{equation}
\int_{\mathcal{F}}\frac{d^2\tau}{\tau_2^2}(\hat
E_2)^r(c_{-1}e^{-2\pi
i\tau}+c_0+\ldots)=\frac{\pi}{3(r+1)}[c_0-24(r+1)c_{-1}]\ ,
\end{equation}
we get,
\begin{equation}
\Lambda_0=\frac{\pi}{2}(\textrm{Re }T_1)\ .
\end{equation}

\subsubsection{Degenerate orbits}
\label{deggp}

In this case we have to compute the contribution,
\begin{equation}
\Lambda_d=-\frac{\textrm{Re }T_1}{8}
\int_{\mathcal{F}_T}\frac{d^2\tau}{\tau_2^2}\left(\frac{60}{\pi\tau_2}-24+\mathcal{O}(e^{2\pi
i \tau})\right)\sum_{j,p}\textrm{exp}\left[-\frac{\pi\textrm{Re
}T_1}{\tau_2\textrm{Re }U_1}|j+ipU_1|^2\right]
\end{equation}
where the integration region, $\mathcal{F}_T$, corresponds to the
upper band $\{|\tau_1|<1/2,\ \tau_2>0\}$. This can be done using
the formula \cite{elias1},
\begin{equation}
\int_{\mathcal{F}_T}\frac{d^2\tau}{\tau_2^{1+r}}\sum_{j,p}\textrm{exp}\left[-\frac{\pi\textrm{Re
}T}{\tau_2\textrm{Re }U}|j+ipU|^2\right]=\frac{2\Gamma(r)\zeta
(2r)}{(\pi \textrm{Re }T)^r}E(iU,r)\ .
\end{equation}
Taking into account that,
\begin{equation}
E(iU,1)=-\frac{3}{\pi}(\textrm{log }|\eta(iU)|^4+\textrm{log
}[(\textrm{Re }T)(\textrm{Re }U)]\mu^2)\ + \ \textrm{const.}\ ,
\end{equation}
with $\mu^2$ the infrared regulator and ``const.'' a
renormalization scheme dependent constant which we will disregard
in what follows, we obtain,
\begin{equation}
\Lambda_d=-3(\textrm{log }|\eta(iU_1)|^4+\textrm{log }[(\textrm{Re
}T_1)(\textrm{Re
}U_1)\mu^2])-\frac{\pi}{3}\frac{E(iU_1,2)}{T_1+\bar T_1}\ .
\end{equation}

\subsubsection{Non-degenerate orbits}
\label{nodeggp}

Finally, for non-degenerate orbits we need to compute,
\begin{multline}
\Lambda_{nd}=-\frac{\textrm{Re
}T_1}{4}\int_{\mathbb{C}^+}\frac{d^2\tau}{\tau_2^2}\sum_{k>j\geq
0,\ p\neq
0}\sum_n\left(d_1(n)-\frac{d_2(n)}{4\pi\tau_2}\right)\ \times \\
\times \ e^{2\pi i \tau n} \textrm{exp}\left[-2\pi
T_1kp-\frac{\pi\textrm{Re }T_1}{\tau_2\textrm{Re
}U_1}|-j-iU_1p+k\tau|^2\right]\ ,
\end{multline}
where we have expanded,
\begin{equation}
\mathcal{A}_f=\sum_n d_1(n)e^{2\pi i n \tau}\ , \qquad
\frac{E_{10}}{\eta^{24}}=\sum_nd_2(n) e^{2\pi i n \tau}
\end{equation}
Performing first the integration on $\tau_1$,
\begin{multline*}
\Lambda_{nd}=-\frac{[(\textrm{Re }T_1)(\textrm{Re
}U_1)]^{1/2}}{4}\sum_{k>j\geq 0,\ p\neq
0}\sum_n\int_0^\infty\frac{d\tau_2}{\tau_2^{3/2}}\frac{1}{k}\left(d_1(n)-\frac{d_2(n)}{4\pi\tau_2}\right)\times
\\
\times \ \textrm{exp}\left[-2\pi i(\textrm{Im }T_1)kp+2\pi i
n\left(\frac{j-p(\textrm{Im }U_1)}{k}\right)\right]\\
\times \ \textrm{exp}\left[-\frac{\pi (\textrm{Re
}T_1)}{\textrm{Re }U_1}\left(k+\frac{n(\textrm{Re
}U_1)}{k(\textrm{Re }T_1)}\right)^2\tau_2-\frac{\pi p^2(\textrm{Re
}T_1)(\textrm{Re }U_1)}{\tau_2}\right]
\end{multline*}
Then the integral on $\tau_2$ can be carried out with the aid of,
\begin{align}
&\int_0^\infty
\frac{dx}{x^{3/2}}e^{-ax-b/x}=\sqrt{\frac{\pi}{b}}e^{-2\sqrt{ab}}\
, \\
&\int_0^\infty
\frac{dx}{x^{5/2}}e^{-ax-b/x}=\left(\frac{1}{2b}+\sqrt{\frac{a}{b}}\right)\sqrt{\frac{\pi}{b}}e^{-2\sqrt{ab}}\
.
\end{align}
And summing over $n$, we finally get,
\begin{equation}
\Lambda_{nd}=-\frac{1}{4}\sum_{k>j\geq 0,\ p>0}\frac{e^{-2\pi
kpT_1}}{kp}\left(\hat{\mathcal{A}}_f(\mathcal{U})+\frac{1}{\pi kp
(T_1+\bar T_1)}\hat{\mathcal{A}}_K(\mathcal{U})\right)\ + \
\textrm{c.c.}
\end{equation}
with $\hat{\mathcal{A}}_f$, $\hat{\mathcal{A}}_K$ and
$\mathcal{U}$ defined in (\ref{elip1}), (\ref{elip2}) and
(\ref{uinst}), respectively.

\subsection{The $SO(q)\times SO(32-q)$ model}
\label{ap2}

\subsubsection{Elliptic genera}
\label{elipo}

The relevant characters for this model are,
\begin{equation}
\chi_o+\chi_v=\frac{E_4^2}{\eta^{16}}\ , \qquad
\chi_o-\chi_v=\frac{1}{2\eta^{16}}(\vartheta_3^{q/2}\vartheta_4^{16-q/2}+\vartheta_4^{q/2}\vartheta_3^{16-q/2})
\end{equation}
Then, it is possible to show that,
\begin{align}
\hat{\mathcal{A}}_{f,1}^{[0,1]}\equiv
&-\frac{\vartheta_3^2\vartheta_4^2}{4\pi\eta^8}\left(\frac{1}{\tau_2}+\frac{\partial^2_{\nu_1}}{\pi}\right)(\chi_o+\chi_v)=\frac{\vartheta_3^2\vartheta_4^2E_4(\hat
E_2E_4-E_6)}{12\eta^{24}}=60-\frac{124}{\pi\tau_2}+\ldots\label{aa1}\\
\hat{\mathcal{A}}_{f,2}^{[0,1]}\equiv \label{aa2}
&\nonumber -\frac{\vartheta_3^2\vartheta_4^2}{4\pi\eta^8}\left(\frac{1}{\tau_2}+\frac{\partial^2_{\nu_1}}{\pi}\right)(\chi_o-\chi_v)=\\
&\qquad
\frac{\vartheta_3^2\vartheta_4^2}{24\eta^{24}}\left[\vartheta_3^{q/2}\vartheta_4^{16-q/2}(\hat
E_2+\vartheta_2^4-\vartheta_4^4)+\vartheta_4^{q/2}\vartheta_3^{16-q/2}(\hat
E_2-\vartheta_2^4-\vartheta_3^4)\right]=\\
&\qquad\qquad\qquad\qquad
=4(q-17)-\frac{q^2-32q-248}{2\pi\tau_2}+\ldots\ .\nonumber
\end{align}
The other modular forms that we need are,
\begin{align}
&\hat{\mathcal{A}}_{K,1}^{[0,1]}\equiv
\frac{1}{4\pi}\left(i\partial_\tau-\frac{1}{\tau_2}\right)\frac{(E_4\vartheta_3\vartheta_4)^2}{\eta^{24}}=
\frac{E_4\vartheta_3^2\vartheta_4^2}{24\eta^{24}}[8E_6+E_4(\vartheta_4^4+\vartheta_3^4)+2\hat
E_2E_4]\\
&\hat{\mathcal{A}}_{K,2}^{[0,1]}\equiv
\frac{1}{4\pi}\left(i\partial_\tau-\frac{1}{\tau_2}\right)\frac{\vartheta_3^2\vartheta_4^2}{2\eta^{24}}(\vartheta_3^{q/2}\vartheta_4^{16-q/2}+\vartheta_4^{q/2}\vartheta_3^{16-q/2})=\nonumber\\
&\qquad
=\frac{\vartheta_3^2\vartheta_4^2}{96\eta^{24}}\left[\vartheta_4^{16-q/2}\vartheta_3^{q/2}\left(8\hat
E_2+\left(14-\frac{3q}{2}\right)\vartheta_2^4+20\vartheta_3^4\right)+\right.\nonumber\\
&\qquad \qquad \qquad \qquad \qquad
\left.+\vartheta_3^{16-q/2}\vartheta_4^{q/2}\left(8\hat
E_2-\left(14-\frac{3q}{2}\right)\vartheta_2^4+20\vartheta_4^4\right)\right]\label{aa4}\
,
\end{align}
Finally, we define $S$ and the $ST^{-1}$ transformed forms,
\begin{align}
\hat{\mathcal{A}}_{f,i}^{[1,0]}(\tau)&=\hat{\mathcal{A}}_{f,i}^{[0,1]}(S\tau)\
, & \hat{\mathcal{A}}_{f,i}^{[1,1]}(\tau)&=
\hat{\mathcal{A}}_{f,i}^{[0,1]}(ST^{-1}\tau)\ , \\
\hat{\mathcal{A}}_{K,i}^{[1,0]}(\tau)&=\hat{\mathcal{A}}_{K,i}^{[0,1]}(S\tau)\
, &
\hat{\mathcal{A}}_{K,i}^{[1,1]}(\tau)&=\hat{\mathcal{A}}_{K,i}^{[0,1]}(ST^{-1}\tau)
\ .
\end{align}
for $i=1,2$.

\subsubsection{Degenerate orbits}

For the degenerate orbits we have the contribution,
\begin{multline*}
\Lambda_{d}=-\frac{1}{8}\int_{\mathcal{F}_T}\frac{d^2\tau}{\tau_2^2}\sum_{p,j}
\{ \left(60-\frac{124}{\pi\tau_2}+\ldots\right)(\textrm{Re
}T_1)\textrm{exp}\left[-\frac{\pi \textrm{Re
}T_1}{\tau_2\textrm{Re
}U_1}\left|j+\frac12+ipU_1\right|^2\right]+\\
\sum_{r=2,3}\left(4(q-17)-\frac{q^2-32q-248}{2\pi\tau_2}+\ldots\right)(\textrm{Re
}T_r)\textrm{exp}\left[-\frac{\pi \textrm{Re
}T_r}{\tau_2\textrm{Re
}U_r}\left|j+\frac12+ipU_r\right|^2\right]\}\ ,
\end{multline*}
where the dots correspond to order $\mathcal{O}(e^{2\pi i\tau})$
terms not contributing to the final expression. Proceeding as in
section \ref{deggp}, we obtain,
\begin{multline}
\Lambda_{d}=-\frac{\pi}{2}\left[5E_{1/2}(iU_1,1)+\frac{q-17}{3}\sum_{r=2,3}E_{1/2}(iU_r,1)\right]+\\
+\frac{\pi}{360}\left[124\frac{E_{1/2}(iU_1,2)}{T_1+\bar
T_1}+\frac12\sum_{r=2,3}(q^2-32q+248)\frac{E_{1/2}(iU_r,2)}{T_r+\bar
T_r}\right]\ ,
\end{multline}
with $E_{1/2}(U,k)$ the shifted non-holomorphic Eisenstein series
defined in (\ref{shifted}).

\subsubsection{Non-degenerate orbits}

We begin computing the contribution from non-degenerate orbits of
type I. This is given by,
\begin{multline*}
\Lambda_{nd_I}=-\frac{1}{4}\int_{\mathbb{C}^+}\frac{d^2\tau}{\tau_2^2}\sum_{p\neq
0, k>j\geq 0}\sum_{n=0}^\infty e^{2\pi i\tau n}(-1)^k\\
\left[ (\textrm{Re
}T_1)\left(d_1(n)-\frac{d_2(n)}{4\pi\tau_2}\right)\textrm{exp}\left(-2\pi
kpT_1-\frac{\pi \textrm{Re }T_1}{\tau_2\textrm{Re
}U_1}|k\tau-j-\frac12-ipU_1|^2\right)+\right.\\
\left. \sum_{i=2,3}(\textrm{Re
}T_i)\left(d_3(n)-\frac{d_4(n)}{4\pi\tau_2}\right)(-1)^{\frac{qk}{4}}\textrm{exp}\left(-2\pi
kpT_i-\frac{\pi \textrm{Re }T_i}{\tau_2\textrm{Re
}U_i}|k\tau-j-\frac12-ipU_i|^2\right)\right]
\end{multline*}
where we have performed the expansions,
\begin{align*}
\mathcal{A}_{f,1}^{[0,1]}&=\sum_{n}d_1(n)e^{2\pi i\tau n}\ , &\frac{(E_4\vartheta_3\vartheta_4)^2}{\eta^{24}}&=\sum_{n}d_2(n)e^{2\pi i\tau n}\ ,\\
\mathcal{A}_{f,2}^{[0,1]}&=\sum_{n}d_3(n)e^{2\pi i\tau n}\ , &
\frac{(\vartheta_3\vartheta_4)^2}{2\eta^{24}}(\vartheta_3^{q/2}\vartheta_4^{16-q/2}+\vartheta_4^{q/2}\vartheta_3^{16-q/2})&=\sum_{n}d_4(n)e^{2\pi
i\tau n}\ .
\end{align*}
Proceeding exactly in the same way as in section \ref{nodeggp}, we
obtain,
\begin{multline}
\Lambda_{nd_I}=-\frac{1}{4}\sum_{p>0, k>j\geq 0}\frac{(-1)^k}{pk}
 \left[ e^{-2\pi pkT_1}\left(\hat{\mathcal{A}}_{f,1}^{[0,1]}(\mathcal{U}_1^{[0,1]})+\frac{\hat{\mathcal{A}}_{K,1}^{[0,1]}(\mathcal{U}_1^{[0,1]})}{\pi kp(T_1+\bar T_1)}\right)\right. \\
\left. +\sum_{r=2,3}(-1)^{\frac{kq}{4}}e^{-2\pi
rkT_r}\left(\hat{\mathcal{A}}_{f,2}^{[0,1]}(\mathcal{U}_r^{[0,1]})+\frac{\hat{\mathcal{A}}_{K,2}^{[0,1]}(\mathcal{U}_r^{[0,1]})}{\pi
kp(T_r+\bar T_r)}\right)\right]\ + \ \textrm{c.c.}\ ,
\end{multline}
with $\mathcal{U}_r^{[h,g]}$ defined in (\ref{udisc}).

For type II (type III) non-degenerate orbits we proceed in the
same way, but performing a change of variables by the
corresponding coset representative, $\tau\to S\tau$ ($\tau\to
ST^{-1}\tau$), obtaining,
\begin{multline}
\Lambda_{nd_{II}}=-\frac{1}{4}\sum_{p>0, k>j\geq
0}\frac{(-1)^j}{p\left(k+\frac12\right)}
 \left[ e^{-2\pi p\left(k+\frac12\right)T_1}\left(\hat{\mathcal{A}}_{f,1}^{[1,0]}(\mathcal{U}_1^{[1,0]})+\frac{\hat{\mathcal{A}}_{K,1}^{[1,0]}(\mathcal{U}_1^{[1,0]})}{\pi p\left(k+\frac12\right)(T_1+\bar T_1)}\right)\right. \\
\left. +\sum_{r=2,3}(-1)^{\frac{jq}{4}}e^{-2\pi
r\left(k+\frac12\right)T_r}\left(\hat{\mathcal{A}}_{f,2}^{[1,0]}(\mathcal{U}_r^{[1,0]})+\frac{\hat{\mathcal{A}}_{K,2}^{[1,0]}(\mathcal{U}_r^{[1,0]})}{\pi
p\left(k+\frac12\right)(T_r+\bar T_r)}\right)\right]\ + \
\textrm{c.c.}\ ,
\end{multline}
and
\begin{multline}
\Lambda_{nd_{III}}=-\frac{1}{4}\sum_{p>0, k>j\geq
0}\frac{(-1)^{k+j}}{p\left(k+\frac12\right)}
 \left[ e^{-2\pi p\left(k+\frac12\right)T_1}\left(\hat{\mathcal{A}}_{f,1}^{[1,1]}(\mathcal{U}_1^{[1,1]})+\frac{\hat{\mathcal{A}}_{K,1}^{[1,1]}(\mathcal{U}_1^{[1,1]})}{\pi p\left(k+\frac12\right)(T_1+\bar T_1)}\right)\right. \\
\left. +\sum_{r=2,3}(-1)^{\frac{(k+j)q}{4}}e^{-2\pi
r\left(k+\frac12\right)T_r}\left(\hat{\mathcal{A}}_{f,2}^{[1,1]}(\mathcal{U}_r^{[1,1]})+\frac{\hat{\mathcal{A}}_{K,2}^{[1,1]}(\mathcal{U}_r^{[1,1]})}{\pi
p\left(k+\frac12\right)(T_r+\bar T_r)}\right)\right]\ + \
\textrm{c.c.}\ .
\end{multline}

\end{document}